# The Current Status of Prime Focus Instrument of Subaru Prime Focus Spectrograph

Shiang-Yu Wang*[a], Mark A. Schwochert[b], Pin-Jie Huang[a], Hsin-Yo Chen[a], Masahiko Kimura[a], Richard C. Y. Chou[a], Yin-Chang, Chang[a], Yen-Sang Hu[a], Hung-Hsu Ling[a], Chaz N. Morantz[b], Dan J. Reiley[c], Peter Mao[c], David F. Braun[b], Chih-Yi Wen[a], Chi-Hung Yan[a], Jennifer Karr[a], James E. Gunn[d], Graham Murray[e], Naoyuki Tamura[f], Naruhisa Takato[g], Atsushi Shimono[f], Decio Ferreira[h], Leandro Henrique dos Santos[h], Ligia Souza Oliveira[h], Antonio Cesar de Oliveira[h] and Lucas Souza Marrara[h]

[a]Institute of Astronomy and Astrophysics, Academia Sinica, P. O. Box 23-141, Taipei, Taiwan;
[b]Jet Propulsion Laboratory, 4800 Oak Grove Dr., Pasadena, CA 91109, USA;
[c]California Institute of Technology, 1200 E California Blvd, Pasadena, CA 91125, USA;
[d]Princeton University, Princeton, New Jersey, 08544, USA;
[e]Durham University, Durham, DH1 3LE, UK;
[f]Kavli Institute for the Physics and Mathematics of the Universe (WPI), The University of Tokyo, 5-1-5 Kashiwanoha Kashiwa Chiba, 277-8583, Japan;
[g]Subaru Telescope, National Astronomical Observatory of Japan, 650 North Aohoku Place, Hilo, Hawaii, USA;
[h]Laboratório Nacional de Astrofísica, Itajubá, 37504-364 Minas Gerais, Brazil;

## ABSTRACT

The Prime Focus Spectrograph (PFS) is a new optical/near-infrared multi-fiber spectrograph design for the prime focus of the 8.2m Subaru telescope. PFS will cover 1.3 degree diameter field with 2394 fibers to complement the imaging capability of Hyper SuprimeCam (HSC). The prime focus unit of PFS called Prime Focus Instrument (PFI) provides the interface with the top structure of Subaru telescope and also accommodates the optical bench in which Cobra fiber positioners are located. In addition, the acquisition and guiding cameras (AGCs), the optical fiber positioner system, the cable wrapper, the fiducial fibers, illuminator, and viewer, the field element, and the telemetry system are located inside the PFI. The mechanical structure of the PFI was designed with special care such that its deflections sufficiently match those of the HSC's Wide Field Corrector (WFC) so the fibers will stay on targets over the course of the observations within the required accuracy. In this report, the latest status of PFI development will be given including the performance of PFI components, the setup and performance of the integration and testing equipment.

## 1. INTRODUCTION

The PFS[1] is a new multi-fiber spectrograph on Subaru telescope. PFS will provide low to medium resolution spectrum for the scientific objects from 0.38µm to 1.26µm. PFS shares the same WFC with HSC[2] which is a new wide field camera with 1.5 degrees field of view. The 2394 fibers populate in a hexagon shape on the prime focal plane of Subaru telescope covering 1.3 degrees diameter field. Each fiber is designed to be driven by a Cobra positioner which has two miniature motors to provide two degrees of freedom in a 9.5 mm diameter patrol region on the focal plane. The PFI is the prime focus unit of PFS to be installed in the prime focus structure called POpt2 of Subaru telescope. The primary functions of PFI is to provide the mechanical interface for the Cobra optical bench (COB) where the science fiber positioners and fixed fiducial fibers are mounted and support the science fiber routing from the focal plane to the spectrographs off the telescope. The PFI includes the AGCs, center viewing camera, cable wrapper, field element, fiducial fiber illuminators, control electronics, and telemetry system inside the PFI. The flat fielding and spectrum calibration lamps are positioned on the top of PFI.

* sywang@asiaa.sinica.edu.tw; phone 886 2 2366-5338; fax 886 2 2367-7849; www.asiaa.sinica.edu.tw

Being at the prime focus environment of Subaru telescope, tight space, weight and heat dissipation constraints are applied to PFI. Furthermore, the structure stiffness of PFI is also limited to avoid any possible damage of the fragile ceramic lens barrel of the wide field corrector. A combination of different materials at different locations in height is used in PFI to provide a stable focal plane position over the operation temperature range (5 to -5 degree C) of PFS. A glycol based cooling system removes the heat generated from the electronics of various components of PFI to avoid possible seeing degradation. To share the same wide field corrector with HSC, a flat glass called field element is added in front of the fibers for compensating the optical path difference.

Six FLI ML4720 cameras are installed at the periphery of the effective field for field acquisition and guiding. A 0.9mm BK7 glass is installed inside each camera to cover half of the CCD sensor to accelerate the focus sequence. Together with a tiny 0.75 mm viewing camera at the field center, AG cameras provide essential information for the pointing and distortion map of the focal plane. The relative distance between the AG camera sensors to the reference fiducial fibers will be calibrated and kept stable within 5 microns with different operation conditions to meet the alignment requirement of the fibers. A high brightness red LED with a light diffusor is used as the light source to illuminate 96 fiducial fibers.

The PFS collaboration is led by Kavli Institute for the Physics and Mathematics of the Universe, the University of Tokyo with international partners consisting of Academia Sinica, Institute of Astronomy and Astrophysics in Taiwan, Caltech/Jet Propulsion Laboratory, Princeton University/John Hopkins University in USA, Chinese PFS Participating Consortium in China, Laboratoire d'Astrophysique de Marseille in France, Max Planck Institute for Astronomy in Germany, National Astronomical Observatory of Japan/Subaru Telescope, and Universidade de São Paulo/Laboratório Nacional de Astrofísica in Brazil.

## 2. PFI COMPONENTS UPDATES

The basic design requirements and the details of PFI components were presented in 2014 SPIE meeting[3]. The PFI passed the critical design review (CDR) in March 2015. Most components follow the original design and have been manufactured accordingly. Figure 1 shows the schematics of the PFI with major components indicated. The major changes after CDR includes the addition of a flat glass in each AGC, the field center camera and the calibration lamp on top of PFI. The mechanical structure has been completed and tested. The following sections describe the updates.

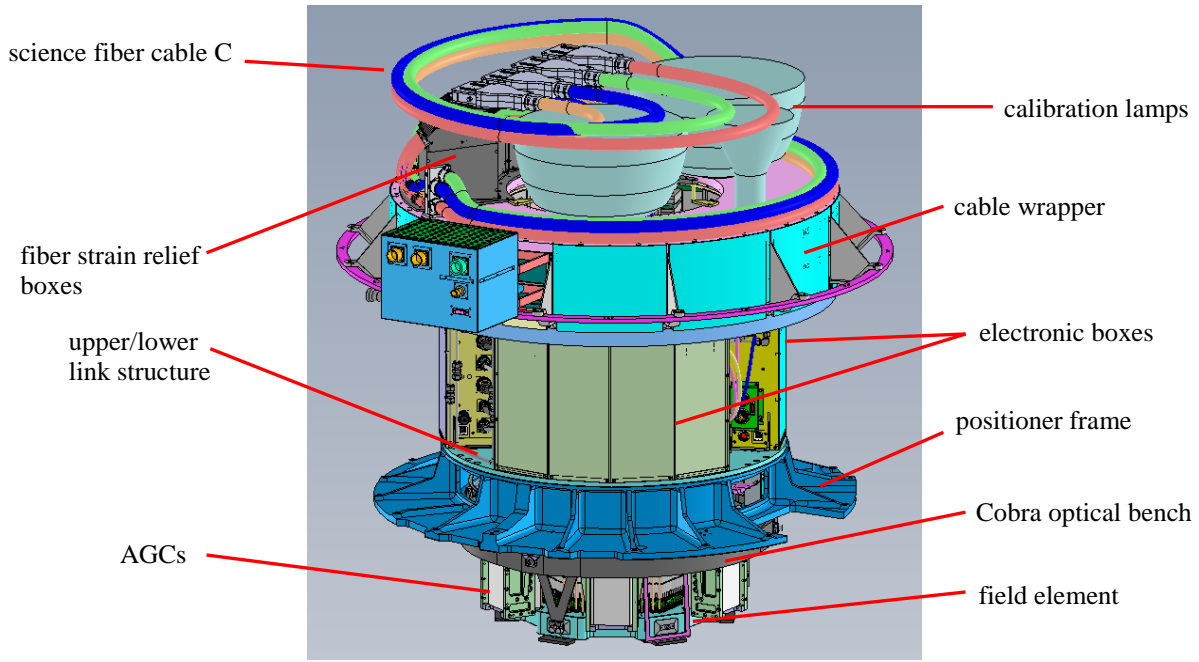

Figure 1. The components of PFI. The top cover for the storage of optical fibers was not shown.

## 2.1 The interface fitting test

After the major structure was delivered, a fitting test was executed at Subaru telescope to verify the interface and the installation process between PFI and POpt2. The positioner frame and the cable wrapper of PFI are the two structures which interface with POpt2. The positioner frame is bolted on the instrument rotator and it has a tight tolerance in size of 66 μm respected to the 1.108m diameter. The tight size tolerance ensures the installation repeatability of PFI to the telescope optical axis. After the delivery, the dimensions of the positioner frame were measured with CMM in an environment with temperature variation less than 0.5 degree C. The measured size was confirmed to be 1107.950 mm which fulfills the specifications. For the cable wrapper, it is mounted at the fixed top surface of POpt2. The tolerance is relatively relaxed since there is a buffer between the fixed part and rotating part of PFI. The fitting test has two main goals: the first is to confirm the interface and the second is to examine the installation and handling procedure of PFI in and out of POpt2. Through the installation practice, the overall size of PFI was also verified to satisfy the space constrains for transportation at the observatory.

Since most PFI components were not ready during the tests, dummy weight blocks were used to make up the weight of components which were not yet delivered. Figure 2 shows the picture of the structure used in the fitting test including the lifting jig. Most of the practices went well during the tests. However, the height of the temporal transportation cart was not correct so we could not install the whole PFI at once. The installation was conducted with two pieces. The cable wrapper was installed separately after the PFI was inside POpt2. The maximum gap between the positioner frame and the POpt2 gear support was measured to be 250 μm by pushing the positioner frame towards one side in POpt2. The 250 μm seemed to be large but it is within the expected range if we consider carefully the CTE difference between the materials used in POpt2. The gap is similar to the case of HSC and the installation is guided with four pins so we expect similar installation repeatability (~50 μm) of PFI can be reached during the operation.

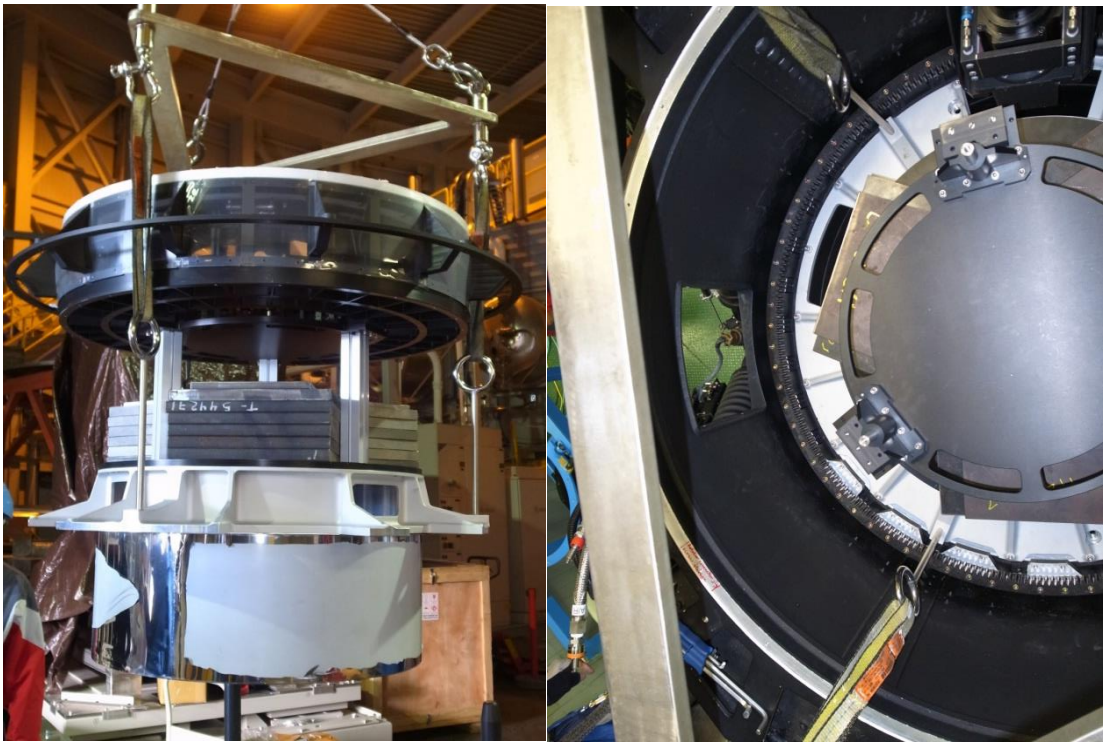

Figure 2. The structure used in the PFI fitting tests (left) and PFI installed inside POpt2 without the cable wrapper (right).

Several issues were discovered during the installation process. Some design changes of the lifting structure and transportation cart have been made to mitigate the issues. The structure and location of the connector box to the POpt2 was also fixed during the test. We are confident that the PFI structure can be smoothly installed into POpt2 with good repeatability after the improvement.

## 2.2 Acquisition and Guiding cameras

The PFI has six FLI ML 4720 frame transfer cameras as AGCs installed at the periphery of the focal plane. AGCs are essential to the focus, field acquisition and guiding of PFI. In order to accelerate the focus process, a flat glass will be added to cover half of the CCD 47-20 sensitive area. The optical path difference between the area with and without the glass generates different image sizes at two regions of the sensor. The AGCs will be mounted so that the images at the two sides of the CCDs have the same radius of point spreading functions (PSF) when PFI is at the best focus. When AGC is out of focus outward, the image size for the area covered with the glass is larger; when AGC is out of focus inwards, the image size for the area without the glass is larger. By comparing sizes of the PSFs from the two sides of CCD, we can make the correct focus adjustment without taking through focus images. The disadvantage of adding the glass is the loss of the sensitive area across the edge of the glass. The light cones pass through the edge of the glass cannot be focused well on the sensor. The loss area is proportional to the distance between the sensor and the glass. After consulting with FLI and e2v, the closest possible distance between the glass and the sensor is 1 mm by utilizing the existing step of the ceramic package of CCD 47-20. Figure 3 shows the CCD package with the glass. This will generate about 0.89mm width of area where the light cones pass through the glass edge. The loss is about 7% of the effective area which is still acceptable.

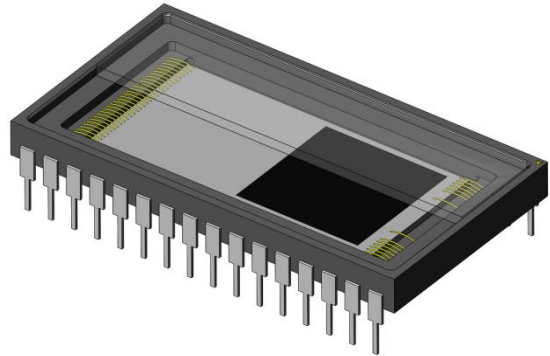

Figure 3. The CCD 47-20 with 0.9mm BK7 glass covering half of the sensor.

In principle, the thickness of the glass should be chosen such that the difference of the PSFs is most sensitive to the focus distance change. However, since the AGC will also provide the guiding signal, the PSFs should be reasonably small when PFI is at focus so that the centroid accuracy and the limiting magnitude of the guide star do not deteriorate too much. The final glass is made of BK7 with a thickness of 0.9mm considering the tradeoffs and the physical limitation of the CCD package. Based on the parameters, the AGC sensor surface will be mounted 0.15mm away from the PFI focal plane. The FWHM of the image will be 12% larger (~78μm) compared with the case if the AGC sensor surface is at the PFI focal plane (~69μm) under 0.7" seeing. In order to make up the signal reduction, AGCs will have a wide passband filter (400-750nm) to receive more photons. Considering all factors, the star counts with the sensors were estimated. For the worst case, when telescope is pointed at galactic pole, 14 stars with SNR larger than 40 will be detected by the six AGCs with 1s exposures on average. Also, with 10s exposures of focus sequences, we will have four stars with SNR larger than 40 with each AGC at galactic pole. We expect to have enough stars with reasonable SNR for focusing and guiding of PFS during the operation. The AGCs are now in FLI for the modification of the glass installation.

One driving requirement for the AGCs is the stability of the positions of the camera sensitive area at different operating temperature and elevation angles. The FLI camera has been tested in an environment controlled chamber and also on a tilt stand with an artificial star and a laser position sensor. The overall image shift on CCD plane from -5 to 5 degree C environment temperature change is roughly 0.17 pixels or 2.18 μm at the center of the sensor. Since the camera sensor is not located at the center of the housing, we observed a larger shift towards the radial direction of the sensor edge. The shift is quite repeatable and can be calibrated with the control software. The height/focus shift of the sensor with the same temperature range is roughly 6 μm. It is small enough for the AGCs to stay in focus. However, the shift in height will generate slight image lateral movement (~0.6 μm). This will also be compensated in the software. For deformation tests, the cameras were mounted on a tilt stand and rotated to simulate the elevation change of the telescope. Since six AGCs are installed on the PFI focal plane with different orientations, the shift of the CCD surface are different for the six cameras at certain elevation angle. The test results showed the largest deformation for all CCDs is about 1.6 μm from 90 to 10 degree elevation change. This is consistent with our structure analysis and also within the required range of deformation.

**2.3 Field center viewing camera**

Since the central part of PFI is populated with fiber motors, AGCs are the only devices to deliver the PFI pointing information in the original design. To reduce the risk during the commissioning process, a small camera will be added at the center of COB. Due to the tight space limits, it is impossible to have a typical science camera installed. Instead, a miniature CMOS camera will be used since only the position at the center of COB needs to be measured. We selected the CMOSIS Naneye camera as the field center camera. It is a 250×250 pixel camera with 3 μm pixels and the physical size of the sensor head is only 1.1 mm. It only requires four wires from its USB controller. The package is small enough so that we can insert the camera and cable into the fiducial fiber tube as shown in figure 4. A special mounting jig is being designed to bond the camera with the fiducial fiber sleeve with a position precision of 0.1mm. The camera has a fixed 44Hz frame rate and the tested quantum efficiency is about 20% at 560nm. With these parameters, the estimated SNR for a magnitude 12 star is roughly 20 for one exposure. By stacking 1 second images, we can reach 14 magnitude star with SNR better than 20. It should not be a problem to find a suitable star for the camera during the commissioning. The position of the camera will be calibrated within 50μm accuracy. Since the FoV of this camera is about 11"×11", it is much larger than the blind pointing accuracy of Subaru telescope plus the PFI installation precision. It will help us to calibrate the center of PFI respected to the telescope pointing at the beginning of the commissioning.

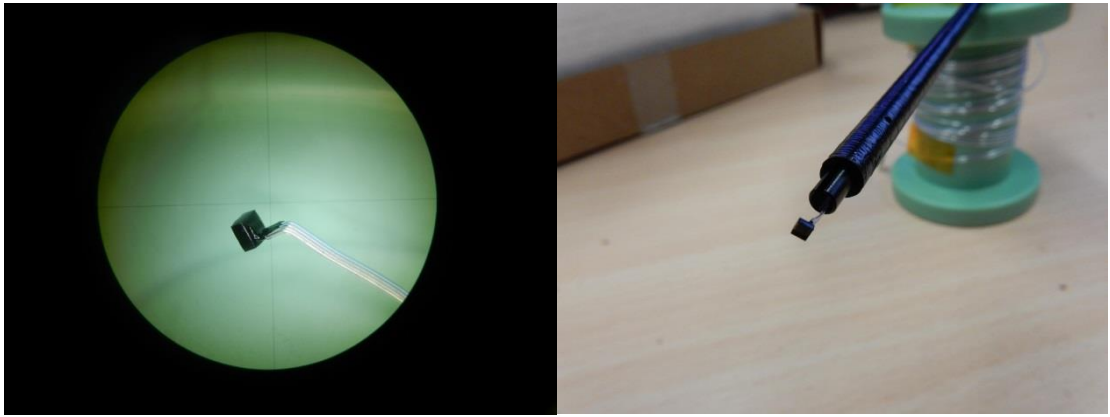

Figure 4. The image of Naneye sensor head under microscope (left) and the device installed in the fiducial fiber structure (right).

## 3. THE INTEGRATION AND TEST PLAN OF PFI

As the components are being delivered, the integration and test (I&T) of PFI started in 2016. In order to better define I&T process, a mockup was built to visualize the space constrains and improve I&T process. The whole I&T process can be divided into three phases. The first phase is I&T for the mechanical structure. This step is to integrate the large components and verify the interfaces are correct and the deformation is within the required range. As mentioned in section 2, the interface with Subaru telescope has been examined. We will integrate the mechanical structure with the Cable wrapper, COB. Dummy weight blocks will also be added to simulate the parts that are not ready for installation. The deformation of COB respected to the POpt2 mounting interface will be measured to confirm the fiber displacement in x, y and z directions is within the requirements under different elevation angles and temperatures. In order to provide the necessary testing conditions, a PFI testing stand and an environmental controlled chamber has been prepared. Figure 5 shows the picture of the test stand and the test chamber. The test stand can tilt the whole PFI from 0 to 90 degrees of elevation angle and rotate the PFI from for 120 degrees to emulate the actual PFI operation condition. The test stand also accommodates a large x-y stage which will be equipped with a fiber characterization system. It can measure the PFI fiber positions in x, y, z directions and tilt angle during the second phase of I&T. The inner dimensions of environmental chamber are 3.5m (L) × 3.5m (W) × 3m (H). The testing stand and PFI will be kept in the environment controlled chamber with a temperature range from room temperature to -10$^o$C during the whole I&T procedure. A temporal metrology camera will be installed on the testing stand to measure the fiber positions as the real PFS metrology camera but with less accuracy. With this camera, we will be able to run the target convergence test at different stage of I&T.

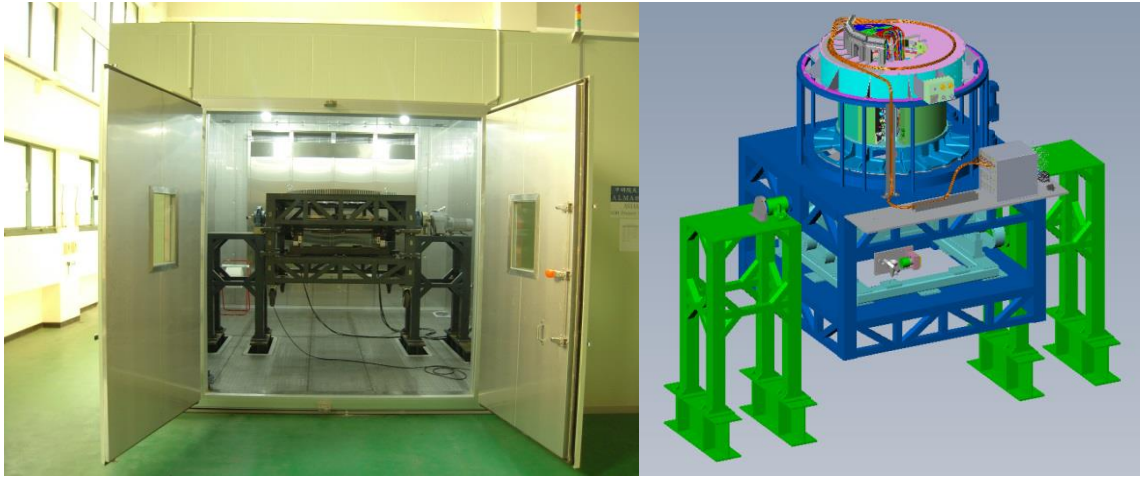

Figure 5. The picture of the testing stand inside the environment controlled chamber (left) and the schematics of the PFI on the testing stand (right).

The second stage of I&T is to install the fibers including the fixed fiducial fibers and the Cobra modules. This is the longest phase for the whole I&T work as each module will be installed and calibrated one by one. Since the PFI focal plane is tightly populated with fibers, the installation sequence for fiducial fibers is very important to avoid difficulties in accessing and replacing components during the installation. The interlaced fiducial fibers will be installed first, followed by the Cobra modules and then the perimeter fiducial fibers. The AGC and its fiducial fibers will be installed as the last components on the focal plane. The Cobra modules will be populated into the three sections on COB evenly; meaning the first three modules will be populated into section 1, 2, and 3 sequentially and then the next three will follow the same procedure. Figure 6 illustrates the installation sequence. The purpose is to reduce the difficulties of removing the Cobra module should any module is damaged during the integration process.

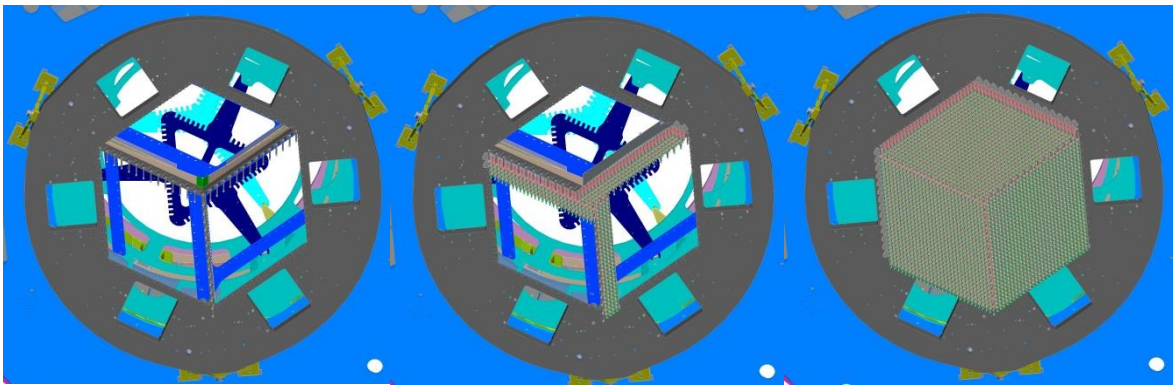

Figure 6. The installation sequence of the Cobra modules. From left to right: COB with interlaced fiducial fibers, first 12 Cobra modules installed, and COB with all Cobra modules installed.

After the installation of each group of fiducial fibers and each Cobra module, the fiber home position on the x-y plane, in the focus direction and the tilt of the fibers will be measured to generate the reference points of the fibers and also confirm the fibers are installed within the allowed position range. Convergence tests will be executed few times throughout the population of the Cobra modules. This is to test the overall control software, speed and algorithm. It is important to find any potential problems before all modules are installed since the installation and removal of the modules are highly risky. The positions of the AGCs will also be calibrated after the installation and adjustment. Based on the required target positioning accuracy and the commissioning plan, the accuracy and range of the measurement items are listed in Table 1. After the focal plane is fully populated, all the measurement will be repeated at different temperatures, elevation angles and rotator angles.

Table 1. The measurement items and accuracy for PFI.

| Type | Target | Reference | Accuracy | Range |
|---|---|---|---|---|
| x/y | • Science fiber home positions<br>• Science fiber home rotation enters<br>• Fiducial fibers<br>• Opaque dots on field element | Fiducial fiber near the field center | +/- 50um | +/- 500um |
| | • AG camera fiducial fibers<br>• AG camera sensors | Nearest perimeter fiducial fibers | +/- 10um | +/- 500um |
| | • Convergence | Requested x/y | +/- 10um | +/- 500um |
| Focus | • Science fibers<br>• Fiducial fibers<br>• AG camera sensors | Designed Focus position | +/- 12um | +/- 500um |
| Tilt | • Science fibers<br>• Fiducial fibers | Fiducial fiber near the field center | +/- 0.17deg | +/- 1.5deg |
| | • AG camera sensors | COB rear surface | +/- 0.17deg | +/- 1.5deg |

Two different measurement systems were designed for the required measurement items. The first system will be used to measure the locations, heights and tilt of all fibers. The second system will be used to measure the position and tilt of AGCs. Both systems can be installed on the x-y stage to scan over the whole focal plane. Figure 7 shows the pictures of the two systems. Both of them are mounted on an Invar base plate to reduce potential dispersion of the measurement at different temperatures. The first system has three different probes to measure the x-y position, the height and the tilt of fibers respectively. A temporal back illumination light source will be connected to the fiber connector for the tests. The light from the fiber will be directed to the three probes with two beam splitters. Each probe has a 1280×1024 CMOS imager with 5.3μm pixels to take the fiber image with reasonable precision. The x-y position probe is an imaging lens set to focus the fiber images on the sensor. The focus probe has a 2×2 Shack-Harmann sensor with the camera positions at the focal plane of the lens array. The fiber movement in height changes the four spot separations equally on the image. The tilt probe is an afocal system with the system pupil locates at the CCD plane. Such design is sensitive to fiber tilt variation while insensitive to fiber lateral shifts. For the AGC test probe, it consists of an artificial star and a Keyence laser position sensor. The artificial star is focused on the AGC to generate a similar spot size of the fiber core and also a well sampled spot with AGC so that the relative position between the AGC and the fiducial fibers can be measured. The laser position sensor is used to measure the distance to the CCD of the AGCs. With the height information at different sensor locations, the tilt of the AGC can also be estimated to an accuracy of 0.009 degrees.

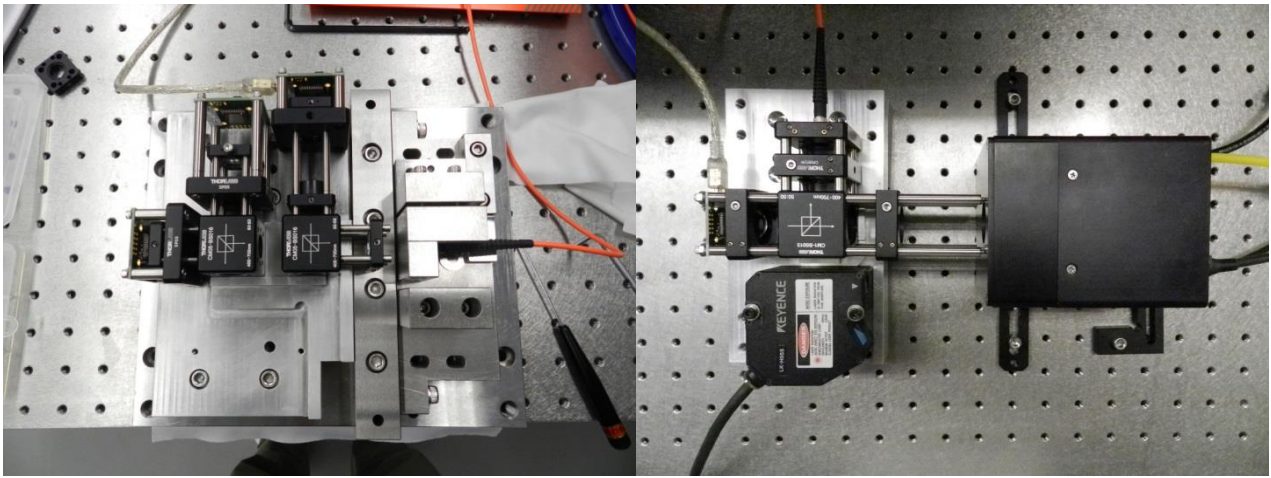

Figure 7. The picture of the fiber test probe (left) and the AGC test probe (right).

The test probes have been calibrated after they were assembled. A temporal fiber with precise positions and tilt were used to simulate the Cobra fibers and to estimate the sensitivity and the possible error of the measurements. The characterization was repeated at different temperatures and different elevation angles to confirm that the measurement precision does not deteriorate with the different environment conditions. Some examples of measured results were plotted in figure 8. For the focus probe, the best linear fit shows a sensitivity of 0.54 μm shift for the spot separation when the fiber height moves for 1 μm. The measurement error is smaller than 0.1 pixels which is less than 1 μm. The plot in figure 8 left panel shows a slightly larger deviation of the four spots when the fiber focus shift is larger than 200 microns. This is due to the fact that the spot becomes quite blurred when the focus offset is large and the measurement accuracy decreases. However, since the fiber focus position accuracy requirement is 50 μm, we can accurately measure the height of the fiber when it is positioned at the right focus range. For the tilt measurement, the image movement slope is 0.069 degree per pixel. This means we can easily measure the tilt change of 0.01 degrees. The position probe shows a linear behavior as expected. The measurements were repeated for several times to examine possible deviation of the measurement. The 1 sigma repeatability of the measurement is 1.2 μm, 0.006 degree and 0.15 μm for the focus, tilt and position probes.

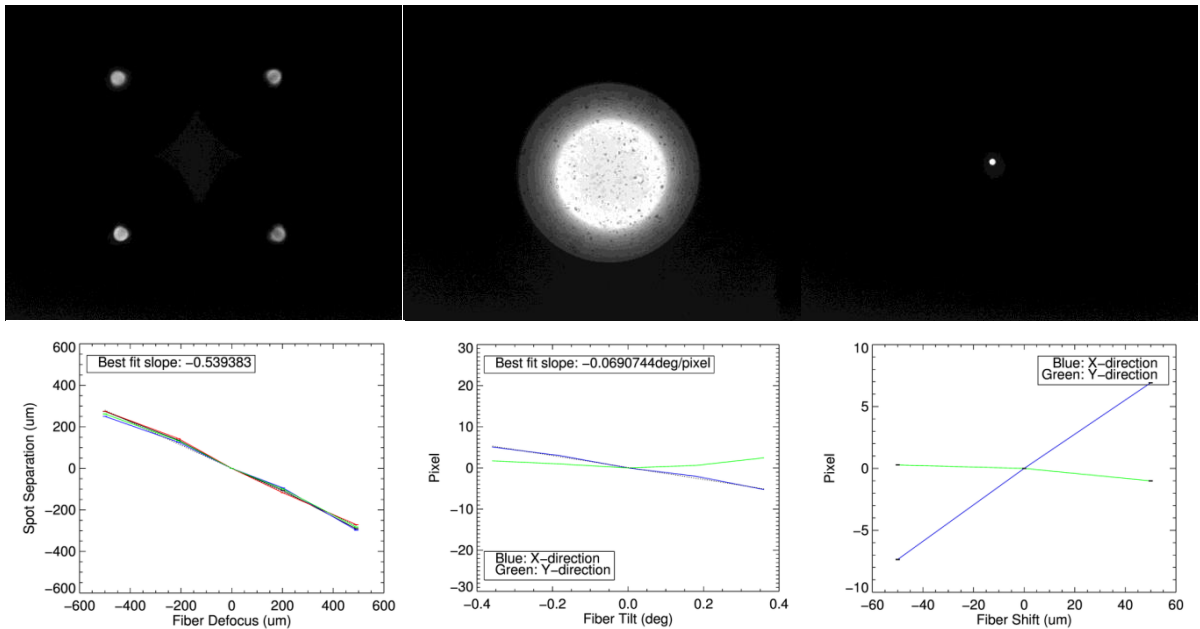

Figure 8. The image of the fiber (top) and the measurement plots (bottom) for the focus, tilt and position probe cameras (from left to the right). The focus probe has 4 image spots so there are 4 lines. The best fitting is the average of the four spots. For the tilt and position measurement, it only shows the calibration in X direction so the deviation in Y is very small.

If the fiber is not at the perfect focus and tilt position, it can generate measurement error of the real fiber position, focus and tilt. The coupling between the measurements were also checked by intentionally offset the fiber to the perfect position. The focus measurement error due to the fiber position shift is about 2 μm and the tilt measurement error is about 0.07 degrees when the fiber is 50 μm off the center of the lens. No detectable difference was found for the fiber focus measurement error from a tilted fiber if the angle is smaller than 0.2 degrees. The error contribution from the defocus to the fiber position measurement is also quite limited. On the other hand, the position measurement can be easily affected by the fiber tilt position. The 0.2 degrees of fiber tilt will generate 8 μm of image shift. All these factors will be calibrated in the measurement software. Combining with the possible measurement error, the overall measurement error is about 1.22 μm, 0.0085 degree and 4.5 μm for the focus, tilt and position probes. All are well within the required measurement accuracy of the PFI fibers.

The calibration was repeated at different environment temperature and gravity directions. We tested the system at 5 and -5 degree C. There is no noticeable change of the measurement result compared with the data at room temperature. Also, the measurement result is not sensitive to the tilt of the system. This ensures the calibration data can be used at different temperature and elevation angle when we measure the fiber positions and tilt.

With a fully populated COB, I&T will enter the final third phase. Four Tower connectors of the science fibers will be made when all Cobras are verified to be within the required accuracy. The field element and electronic boxes will also be installed. The overall system performance will then be verified. The PFI size, weight, power consumption, rotational torque, and power dissipation will be measured. The most important test is the convergence test at different temperatures, elevation angles and rotator angles. The target convergence should be completed within 2 mins and 10 μm positioning accuracy. Once we passed the convergence test, PFI will be packed and delivered to Subaru telescope for commissioning. It is now scheduled in early 2018.

## 4. SUMMARY

The current status of the PFI and I&T plan are presented. The main structure components are ready or close to be delivered. We plan to start the integration in summer 2016. Cobra modules will be delivered from late 2016 with the speed of one module per week. PFI I&T will continue for 18 months in Taiwan. We plan to complete I&T in early 2018 for the PFI to be shipped to Subaru telescope.

## ACKNOWLEDGEMENT

We gratefully acknowledge support from the Funding Program for World-Leading Innovative R&D on Science and Technology(FIRST) "Subaru Measurements of Images and Redshifts (SuMIRe)", CSTP, Japan for PFS project. The work in ASIAA, Taiwan is supported by the Academia Sinica of Taiwan.